# Asymmetric EPR entanglement in continuous variable systems

Katherine Wagner, Jiri Janousek, Seiji Armstrong, Jean-François Morizur, Ping Koy Lam, and Hans-Albert Bachor
*Australian Centre of Excellence for Quantum-Atom Optics*
*Australian National University.*
(Dated: March 9, 2012)

Continuous variable entanglement can be produced in nonlinear systems or via interference of squeezed states. In many of optical systems, such as parametric down conversion or interference of optical squeezed states, production of two perfectly symmetric subsystems is usually used for demonstrating the existence of entanglement. This symmetry simplifies the description of the concept of entanglement. However, asymmetry in entanglement may arise naturally in a real experiment, or be intentionally introduced in a given quantum information protocol. These asymmetries can emerge from having the output beams experience different losses and environmental contamination, or from the availability of non-identical input quantum states in quantum communication protocols. In this paper, we present a visualisation of entanglement using quadrature amplitude plots of the twin beams. We quantitatively discuss the strength of asymmetric entanglement using EPR and inseparability criteria and theoretically show that the optimal beamsplitter ratio for entanglement is dependent on the asymmetries and may not be 50/50. To support this theory, we present experimental results showing one particular asymmetric entanglement where a 0.78/0.22 beamsplitter is optimal for observing entanglement.

## I. INTRODUCTION

The EPR paradox was originally formulated in 1935 as a *Gedanken* experiment regarding quantum mechanics. The argument considers two spatially separated particles that are "entangled" to reveal an inconsistency between the completeness of quantum mechanics and local realism. Since then, the concept of entanglement has transformed from being purely theoretical, to being experimental, and more recently, to being applicable to a range of quantum information applications [1]. Entanglement can now be quantified using various criteria, ranging from violations of Bell's inequalities to demonstrations of inseparability, EPR paradox, entropy of entanglement, and a range of entanglement witnesses [2]. These benchmarks have been extended from discrete bipartite systems to continuous variable as well as multi-partite systems to quantify entanglement specific to its application. For example, violation of Bell's inequalities uses entangled beams to show the implausibility of hidden variable theories. More recently, quantum discord presents an interesting informatic perspective for quantifying quantum correlations in quantum information protocols [3]. In the Schrödinger picture, an entangled state leads to a wavefunction that cannot be described as a product of the two sub-systems, and this gives rise to the inseparability criterion [4]. In the Heisenberg picture, on the other hand, it is the inference of the conjugate variables in one sub-system by measuring the other sub-system that provides a sufficient condition for entanglement, and this is the reasoning behind EPR entanglement.

The originally conceived form of entanglement, the EPR paradox with a pair of particles, can also be applied to continuous variable systems [5]. It involves the measurement of one of the entangled modes to attempt to predict the behaviour of the second mode to a higher degree of accuracy than can be measured by classical means. This results in an apparent violation of the Heisenberg Uncertainty Principle, since this measurement is more precise than allowed for an instantaneous measurement in two conjugate variables.

EPR entanglement has potential applications such as entanglement-based quantum key distribution (QKD)[6], quantum teleportation [7], and entanglement swapping [8]. Observable cases of EPR entanglement have been extended beyond the original two particle case to include entanglement between objects such as spin-polarised atoms [9], Bose-Einstein condensates [10], and laser beams [5, 11], and it is the latter case that we will focus on here.

While normally the equations used assume that the entanglement setup is symmetric, this is rarely the case experimentally. Assuming that the system is symmetric when it is not can result in unoptimised measurements for the entanglement, and in these cases a more rigorous model is required to achieve better results.

This paper is divided into the following sections. In sec. II, we outline the correlations that arise in continuous variable entanglement using elliptical distributions to visualise these correlations. In sec. III, we discuss the asymmetry inherent in the EPR criterion and how this affects the paradox given losses in one of the entangled beams. In sec. IV, we look at the case of asymmetry before the beamsplitter, in particular the case with only one squeezer. The ideal beamsplitter ratio given the loss in the squeezer is calculated, and the result is verified experimentally.

## II. ENTANGLEMENT CORRELATIONS

Entanglement in continuous variable systems manifests itself as correlations between the two entangled objects. These correlations can be used to quantify the entangle-ment on the various scales that have been defined, such as the inseparability criterion, EPR paradox, and so on.

For two quantum noise limited beams that are combined on a beamsplitter as shown in Figure 1, the resulting output beams, while classically identical, are uncorrelated at a quantum level. The instantaneous fluctuations that occur in such a 2 mode system are represented in the traces in Figure 2a. Since these are uncorrelated, the combination of the two traces by addition or subtraction gives the same variance, as does any normalised linear combination of the two beams. If the values of the two measurements are plotted parametrically as previously done in [12], this results in a circular distribution as seen in Figure 2b. Each point in this diagram represents one pair of data points, $\delta_x$ and $\delta_y$, where the data were taken in the same integration time. If we then draw a line showing the standard deviation $\sigma$ of the distribution for all possible angles, the resulting circle defines our quantum noise limit for such a system, as shown in Figure 2c. Measurements of laser beams can be made on the amplitude quadrature, represented by $X^+$, the phase quadrature, represented by $X^-$, or at any phase angle in between. Because in this case the input state has the same noise statistics independent of the quadrature (provided the input states are quantum noise limited), the distribution of the mixed modes looks identical for any given quadrature.

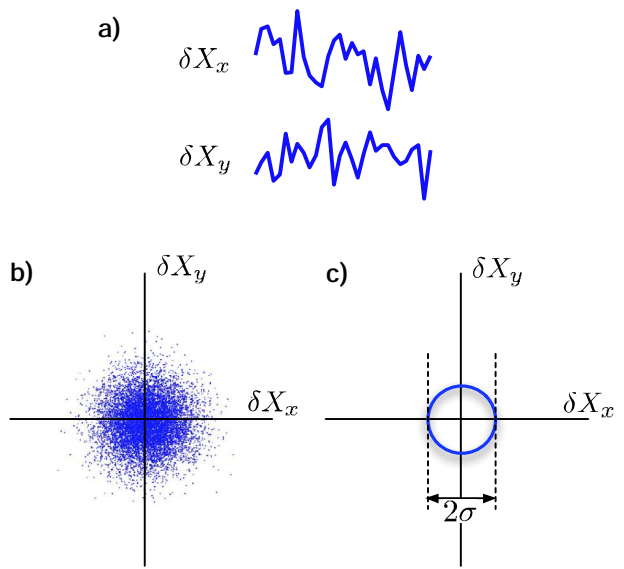

FIG. 2. In the unentangled case with two quantum noise limited beams, the noise on the beams $\delta X_x$ and $\delta X_y$ is uncorrelated, as seen in **a**, resulting in a circular distribution, seen in **b** and **c**. This is true for any given quadrature.

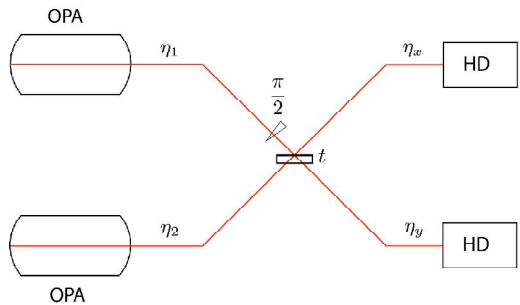

FIG. 1. A standard entanglement experiment, showing where the losses can fall.

If the two input coherent states are replaced with squeezed states, and mixed with a phase difference of $\pi/2$, the two modes, now entangled, are each still noisy, but the noise on each of the modes is now anti-correlated, as seen in Figure 3a. The normalised sum of the two sets of points then fluctuates by less than either of the individual traces, while the difference fluctuates more. This means that if each pair of points is again plotted parametrically, an elliptical shape will result, as shown in Figure 3b and c. Note that, as with squeezing, the Heisenberg Uncertainty Principle is not being violated with the measurement, that falls below the QNL, since making this same measurement on the other quadrature will be above the QNL by at least the same factor as this measurement is under. The system represented by the ellipses shown is assumed to be lossless, and is symmetric, so the two squeezers being mixed are identical.

From the shape of the ellipse that has been obtained, several values useful for categorising the state can be found, as shown in Figure 4. This includes (for a given quadrature) the standard deviation for either of the entangled modes on their own ($\sigma_x$ and $\sigma_y$), and the standard deviations of the sum ($\sigma_{x+y}$ - here lower than the QNL) and difference ($\sigma_{x-y}$) of the modes. From the standard deviations, we can find the variances of the modes from $V = \sigma^2$.

Given that we're now looking at states that aren't coherent, the shape of the distribution changes for different quadratures measured. The phase quadrature will have a correlation between the noise traces, resulting in an ellipse at 90° to that in the amplitude quadrature case. The ellipses for the different quadratures is shown in Figure 5. From each cross section, similar noise statistics can be inferred.

The noise statistics of the entangled state can be visualised using such methods, but it is not useful for characterising the state in a complete and concise way. For this, we will introduce the correlation matrix. Light with Gaussian statistics, as we generally encounter in the continuous variable regime, can be fully characterised by an appropriate correlation matrix. For a system with two modes, $x$ and $y$, this correlation matrix is defined as

$$CM = \begin{bmatrix} C_{xx}^{++} & C_{xx}^{+-} & C_{xy}^{++} & C_{xy}^{+-} \\ C_{xx}^{-+} & C_{xx}^{--} & C_{xy}^{-+} & C_{xy}^{--} \\ C_{yx}^{++} & C_{yx}^{+-} & C_{yy}^{++} & C_{yy}^{+-} \\ C_{yx}^{-+} & C_{yx}^{--} & C_{yy}^{-+} & C_{yy}^{--} \end{bmatrix} \qquad (1)$$

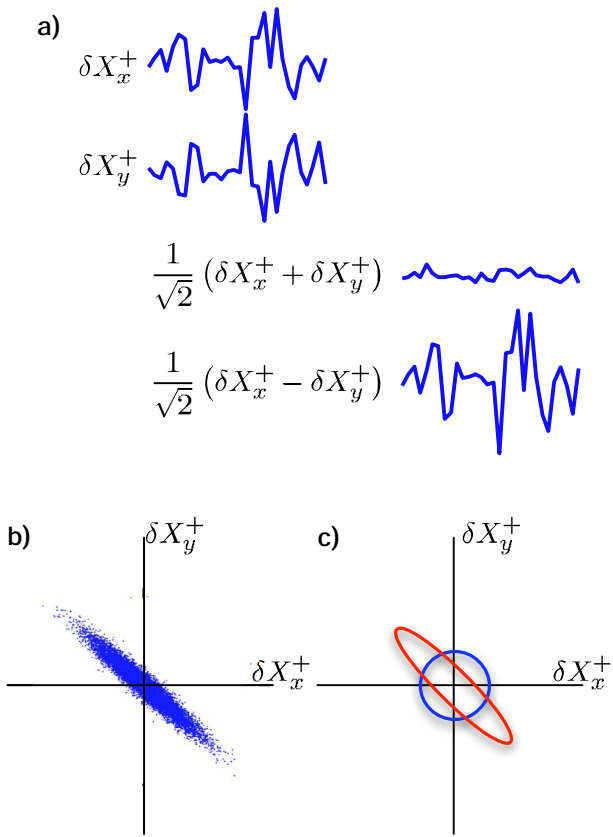

FIG. 3. In this case, $X_x^+$ and $X_y^+$ are correlated to below the QNL, resulting in an elliptical distribution. In part c), the distribution for this case is shown in red, and the quantum noise limit is shown in blue for comparison.

Each element in the matrix is defined as

$$C_{mn}^{kl} = \frac{1}{2}\langle \hat{X}_m^k \hat{X}_n^l + \hat{X}_n^l \hat{X}_m^k \rangle - \langle \hat{X}_m^k \rangle \langle \hat{X}_n^l \rangle \qquad (2)$$

where $\{k, l\} \in \{+, -\}$ and $\{m, n\} \in \{x, y\}$.

Some of these elements correspond directly to the variances that can be found from the ellipses shown above - for instance $C_{xx}^{++}$ is $V_x^+$.

Note that although the matrix uses the amplitude and phase quadratures explicitly, the state can equivalently be characterised using the more general quadratures, $\theta$ and $\theta + \pi/2$.

While the required information about the state is contained in the correlation matrix, it is not immediately obvious by looking at the matrix whether the state is entangled. Entanglement is usually verified with one of several definitions. The two measurements that are discussed here are the inseparability of a system, and the EPR paradox.

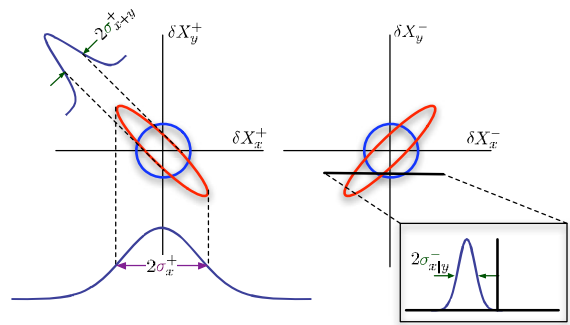

FIG. 4. From the elliptical distributions, we can find $\sigma_x^\pm$, $\sigma_y^\pm$, combinations of these, such as $\sigma_{x+y}^\pm$, and the conditional deviation $\sigma_{x|y}^\pm$.

### Inseparability criterion

For a symmetric case, with identical squeezers and a 50:50 beamsplitter, the end value that is found is given by [13]:

$$\mathcal{I} = \frac{V_{x+y}^\pm + V_{x-y}^\mp}{2} \qquad (3)$$

which can be found directly from two measurements, without the need to find all of the components of the correlation matrix. If $\mathcal{I}$ is found to be less than one, the state is said to be inseparable, and hence entangled. At worst, this is an unoptimised form of inseparability, and it is a sufficient but not a necessary condition for entanglement.

These two measurements required to find the unoptimised inseparability can be found from the correlation ellipses for the phase and amplitude quadrature of an entangled system, as seen in Figure 4.

For a symmetric system, with two squeezed modes with initial amount of squeezing $V_0$, the measurement of inseparability becomes less marked with loss. As a function of the initial amount of squeezing and the transmission $\eta$, the inseparability is given by:

$$\mathcal{I} = \eta V_0^+ + (1 - \eta). \qquad (4)$$

## III. EPR CRITERION AND ASYMMETRY IN THE ENTANGLED BEAMS

Mathematically, the degree of EPR paradox is described in terms of the conditional variance of the system.

$$\epsilon = V_{x|y}^+ V_{x|y}^- \qquad (5)$$

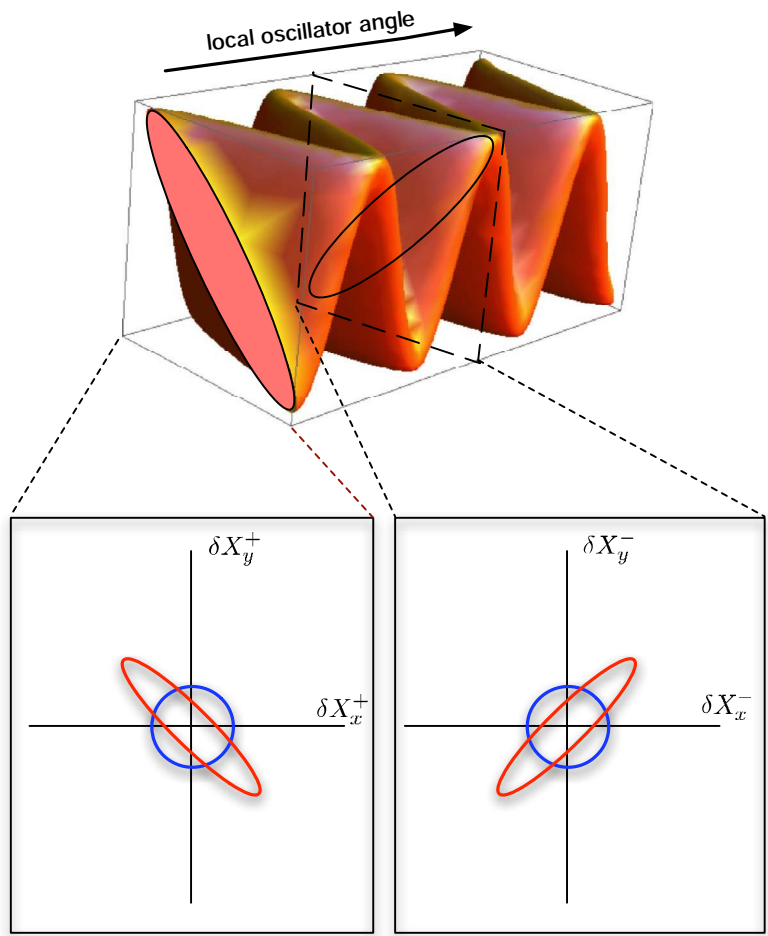

FIG. 5. The correlation ellipses for a continuously varying quadrature (local oscillator angle). The phase and amplitude quadrature cross-sections are shown.

where the conditional variance $V_{x|y}^{\pm}$ denotes the variance of a mode $x$ given a measured value of $y$, and is found by:

$$V_{x|y}^{\pm} = min_{g^{\pm}} \left\langle (\delta X_x^{\pm} - g\delta X_y^{\pm})^2 \right\rangle \quad (6)$$

$$= V_x^{\pm} - \frac{|\langle \delta X_x^{\pm} \delta X_y^{\pm} \rangle|^2}{V_y^{\pm}}. \quad (7)$$

Alternatively, as with the simplified version of the inseparability measure, the EPR measure can be obtained directly from correlation ellipses already discussed. The conditional deviation $\sigma_{x|y}^{\pm}$ can be found from the intercepting points of the ellipse with the $\delta X_x^{\pm}$ axis, as shown in Figure 4. From these conditional deviation values for two orthogonal quadratures, the conditional variances and hence the value for $\epsilon$ can be found.

Unlike inseparability, measurement of the EPR paradox provides a sufficient but not necessary condition for entanglement. For a symmetric system, the degree of EPR paradox can be written in terms of the initial amount of squeezing $V_0^+$, and the total transmission of the system $\eta$, to be:

$$\epsilon = 4\left(1 - \eta + \frac{2\eta - 1}{\eta\left(V_0^+ + 1/V_0^+ - 2\right) + 2}\right)^2 \quad (8)$$

As with the inseparability, a measurement that finds $\epsilon$ to be less than one implies the existence of entanglement. When measuring the EPR value in an entangled system, the direction of the inference that is being made can in some cases be important, as we show later in this study. The EPR value depends on the conditional variance of the system (Eq. 5), and the conditional variance essentially involves the measurement of one beam and the use of this value to infer the measurement that will be made on the other beam. There are two conditional variance measurements that can be made: $V_{x|y}^{\pm}$ and $V_{y|x}^{\pm}$, as defined in Eq. 7. In the former, beam $y$ is measured and used to predict beam $x$, and in the latter, beam $x$ is measured and used to predict beam $y$. These measurements are termed direct reconciliation and reverse reconciliation respectively. The two EPR values are found from these conditional variances, with $\epsilon_{x|y} = V_{x|y}^{+}V_{x|y}^{-}$ and $\epsilon_{y|x} = V_{y|x}^{+}V_{y|x}^{-}$.

Calculations pertaining to the entanglement between continuous variable beams normally assume that there are two identical squeezers that are mixed with a perfect 50:50 beamsplitter and that the two entangled beams also have identical losses. While this is an adequate approximation for many experiments, it places extra restrictions on the resources that are used, and also results in unoptimised entanglement measurements in cases where the asymmetry is non-negligible. Thus a more complete calculation incorporates the asymmetry that occurs in cases such as biased entanglement is beneficial in order to optimise measurements in these cases.

In order to investigate the effects of losses in the system, we must first identify the possible places where losses can occur. Figure 1 shows a standard setup, with input beams 1 and 2, and output beams $x$ and $y$. Here the OPAs and the homodyne detectors are considered to be perfect, with any losses at these elements combined into the generic losses at the appropriate positions. Each beam has a power transmission given by $\eta_1$, $\eta_2$, $\eta_x$, or $\eta_y$, and the loss on each beam is then $1 - \eta$. Note that this simple model assumes that there is no excess noise being coupled into the system.

Figure 6a uses correlation ellipses to show the different measurements that lead to the two possible EPR values. This case has two -7dB squeezers incident on a 50:50 beamsplitter, with no losses in the system. For the ellipses shown here, the two measurements that can be made, $\epsilon_{x|y}$ and $\epsilon_{y|x}$, are identical, due to the symmetry of the system. When there are no losses present in the system, the semi-minor axis on the ellipse for one quadrature is the inverse of the semi-major axis on the ellipse for the other quadrature, when all measurements are normalised to the QNL. This shows that the state, if Gaussian, is minimum uncertainty, and is analogous to the case when the two quadratures for a squeezed state have a product of one. As expected, losses then increase this product.

If we consider the case where an extra loss is applied to one of the entangled arms, but not the other, then we arrive at the ellipses shown in Figure 6b. Here we have the same -7dB squeezers, but there is a 50% loss on the $y$ beam after the beamsplitter. Now the ellipses are wider, and there is no longer symmetry about the diagonal axes, which are marked with broken lines. While the EPR measurement would still show entanglement for $\epsilon_{y|x}$, it can no longer be observed for $\epsilon_{x|y}$. It is clear that the direction of the inference is now a crucial factor in the final number measured for the EPR entanglement. An extra loss on the measured beam is strongly detrimental to the EPR value obtained, but loss on the predicted beam has less of an effect on the entanglement value measured. If we evaluate the two $\epsilon$ values as a function of the transmission $\eta_y$, we find that even as the transmission of the $y$ channel approaches zero, the value for $\epsilon_{y|x}$ is less than one, as shown in Figure 7. Thus EPR entanglement can still be witnessed in a system where one of the entangled arms is very lossy.

EPR entanglement from directional measurement is of interest to Quantum Key Distribution (QKD) systems, where two parties, often termed Alice and Bob, attempt to establish a secure key while in the potential presence of an eavesdropper, Eve. A secure key can be established when there is a net information rate of greater than one [15]. Once a secure key has been established, the key can then be used to send encrypted information between the two parties. Reverse reconciliation is used in cryptography protocols: if Alice is creating the data to establish a key, then her measurement can be expected to have a lower loss than Bob's measurement, since his beam has travelled through a channel with some loss [16]. The two measurements must then be compared either by Bob predicting Alice's measurement (direct reconciliation) or by Alice predicting Bob's measurement (reverse reconciliation). The net information rate is then given by

$$\Delta I = \frac{1}{2} \log_2 \left( \frac{V_{A|E}^{+} V_{A|E}^{-}}{V_{A|B}^{+} V_{A|B}^{-}} \right) \qquad (9)$$

for direct reconciliation or

$$\Delta I = \frac{1}{2} \log_2 \left( \frac{V_{B|E}^{+} V_{B|E}^{-}}{V_{B|A}^{+} V_{B|A}^{-}} \right) \qquad (10)$$

for reverse reconciliation. Reverse reconciliation is the more favourable option, because of the loss experienced on the beam measured by Bob. This is equivalent to using $\epsilon_{y|x}$ instead of $\epsilon_{x|y}$ above. A review of quantum information protocols in the continuous variable regime can be found in Braunstein and van Loock [3].

## IV. BIAS BETWEEN THE INPUT SQUEEZED BEAMS

In all of the cases considered so far, the entanglement has been the result of two similar squeezers whose outputs were mixed with a 50:50 beamsplitter. While this assumption simplifies the description of the output beams, entanglement can also be achieved when there is a significant asymmetry between the input beams, a case known as *biased entanglement* [17]. In some cases, EPR entanglement can still be observed with just one input squeezer. While the entanglement doesn't exhibit the same degree of correlation as a two squeezer setup, the corresponding simplification of the experiment can prove this a worthwhile concession. Apart from negating the necessity of building and locking one extra squeezer, having one vacuum input into the beamsplitter means that the usual need to mode match the beams and lock the phase of the beams is no longer relevant. **Previously, van Loock and Braunstein [14] showed theoretically that N-mode entanglement can be created from a single squeezed vacuum, enabling teleportation.**

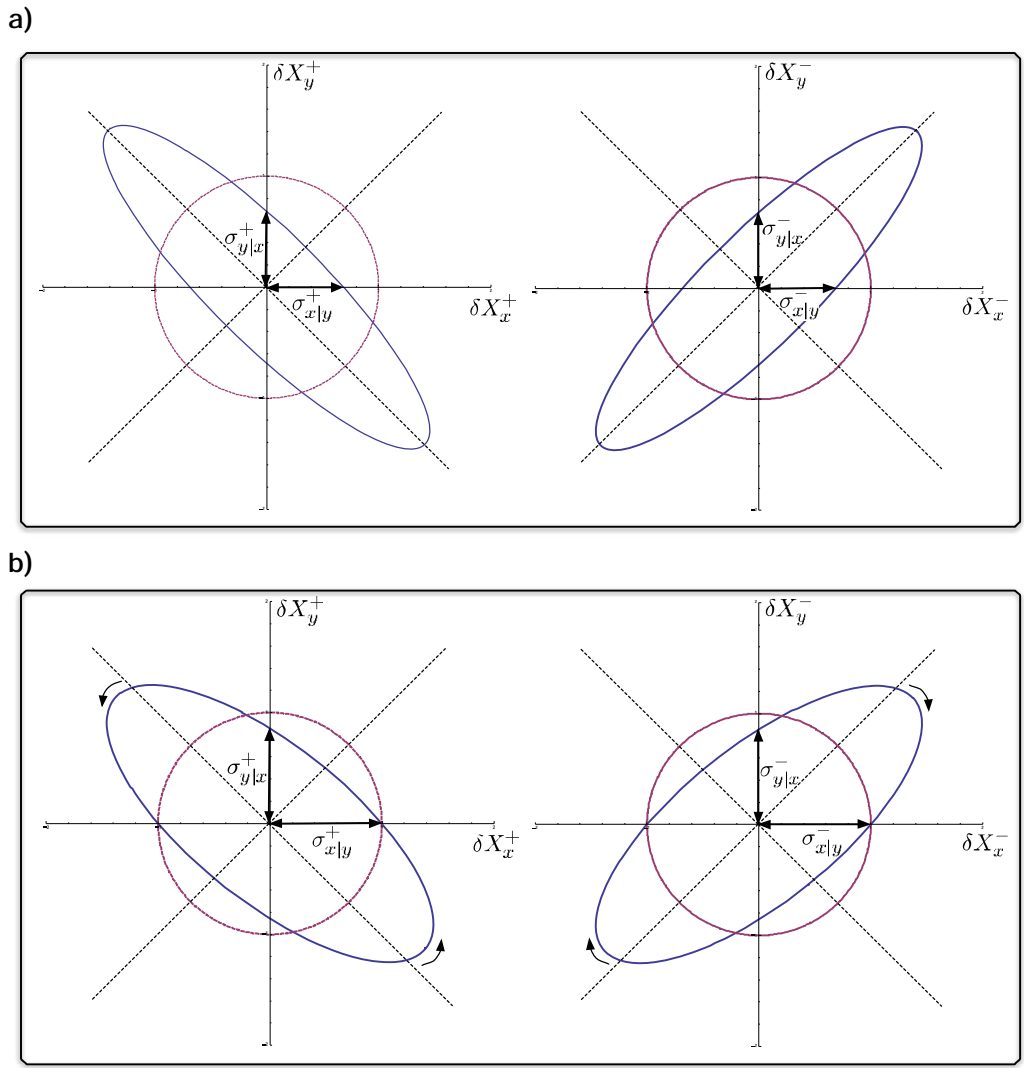

FIG. 6. **a)** Correlation ellipses for a symmetric system, with $\eta_1 = \eta_2 = \eta_x = \eta_y = 1$. The initial amount of squeezing is set to -7 dB. The circles show the quantum noise limit, and the conditional deviation is marked. The conditional variance is the square of this value, $V_{x|y}^\pm = \left(\sigma_{x|y}^\pm\right)^2$. **b)** Correlation ellipses for a system with two -7dB squeezers incident on a beamsplitter. The $y$ beam then undergoes a 50% loss, while the $x$ beam has perfect transmission. EPR entanglement can now be seen with the inference in one direction ($\epsilon_{y|x}$), but not the other ($\epsilon_{x|y}$).

The biased entanglement setup used in [17] consisted of a single squeezer and a 50:50 beamsplitter. This is equivalent to the setup shown in Figure 1 with $\eta_2 = 0$. The two output beams can then be described as:

$$V_x^\pm = V_y^\pm = \frac{1}{2}\left(V_1^\pm + 1\right) \qquad (11)$$

and the EPR measure is found to be:

$$\epsilon_{x|y} = \frac{4V_1^+ V_1^-}{V_1^+ V_1^- + V_1^+ + V_1^- + 1}. \qquad (12)$$

For a biased entanglement setup with a 50:50 beamsplitter, the existence of EPR entanglement can only be measured for a transmission $\eta_x > \frac{2}{3}$. A setup with $\eta_x = \eta_y = 1$ results in the correlation ellipses shown in Figure 8a. One of the axes of each ellipse is coupling in vacuum noise, and so are the same as the quantum noise limit. Adding a loss to the squeezer (by decreasing $\eta_1$ from one) changes one axis on each of the ellipses, and as expected leaves the other unchanged, as seen in Figure 8b.

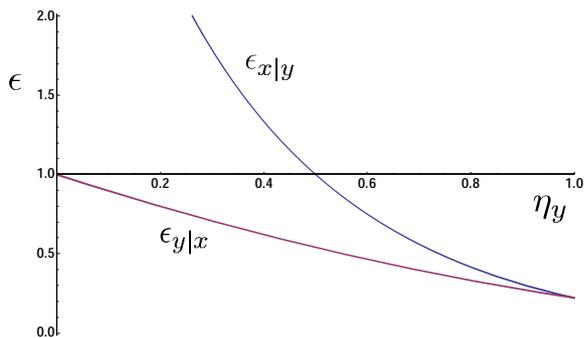

FIG. 7. Measure of EPR entanglement, $\epsilon_{x|y}$ and $\epsilon_{y|x}$ for a system with two -7dB squeezers incident on the beamsplitter. The transmissions $\eta_1$, $\eta_2$ and $\eta_x$ are one, and the transmission $\eta_y$ varies from 0 to 1.

### A. Beamsplitter Ratio

The 50:50 beamsplitter that has featured in experiments until now is the optimal ratio for an symmetric setup, but it can easily be changed should a different ratio be preferable for a given experiment. If we investigate the more general case of an arbitrary beamsplitter ratio in a biased entanglement setup, we instead get two output beams with variance:

$$V_x^\pm = V_1^\pm(1-t) + t$$
$$V_y^\pm = V_1^\pm t + (1-t). \qquad (13)$$

where $t$ is the transmission of the beamsplitter used.

In terms of the correlation ellipses, such a change in beamsplitter ratio corresponds to a rotation of the ellipses from their original positions. The angle of the ellipse axis is then given by $\phi = \frac{\pi}{2}t$, as shown in Figure 9.

Such a rotation can lead to an improvement in the EPR entanglement for one direction, at the expense of the EPR entanglement in the other direction, as seen graphically in Figure 9.

The EPR measurement that can be made for a given squeezer and beamsplitter transmission $t$ is then:

$$\epsilon_{x|y} = \frac{V_1^+ V_1^-}{\left(1 + t\left(V_1^- - 1\right)\right)\left(1 + t\left(V_1^+ - 1\right)\right)}. \qquad (14)$$

While the details here are for the case with just one squeezer, the same optimising principle can be applied to other systems with two squeezers and with different combinations of $\eta_1$, $\eta_2$, $\eta_x$, and $\eta_y$. An introduced asymmetry in the beamsplitter can only lower the EPR value if there is some other asymmetry elsewhere in the experiment. In order to introduce asymmetry and maximise the effect, we have taken the limiting case of using one squeezer out of the many possible experimental realities.

Rather than characterising the OPA by the measured squeezing and antisqueezing, it can be expressed as an initial amount of squeezing $V_0^+$ and a loss term $\eta_1$. Figure 10 shows the EPR value for a varying transmission for the squeezer $\eta_1$ and a varying beamsplitter transmission, $t$. The initial amount of squeezing is constant at -7dB. The section to the right of the EPR entanglement boundary show where the value of $\epsilon_{x|y}$ gives a value of less than one. The position that corresponds to a beamsplitter ratio of 0.5 is marked with the horizontal line. The transmission of the squeezer that was used to test the theoretical results is shown with the vertical broken line.

The cross section from Figure 10 that corresponds to the 50:50 beamsplitter case is shown in Figure 11. For this beamsplitter transmission the efficiency of the squeezer must be $\eta_1 > \frac{2}{3}$. This result is independent of the original amount of squeezing $V_0^+$, and demonstrates one of the main limitations of such a setup.

In the absence of asymmetric losses after the beamsplitter, the ideal transmission can be found by differentiating the EPR value with respect to $t$, and is given by:

$$t_{opt} = \frac{1}{2\eta_1}. \qquad (15)$$

If the beamsplitter is set to the optimum value, then the minimum transmission required on the squeezed beam to observe EPR entanglement decreases from $\eta_1 > \frac{2}{3}$ to $\eta_1 > \frac{1}{2}$. This is shown for the case of a -7dB squeezer in Figure 11. Note that for a symmetric entanglement experiment with two identical squeezers, the transmission on each squeezer must be $\eta > \frac{1}{2}$ in order to measure EPR entanglement. While biased entanglement with a 50:50 beamsplitter requires a high quality squeezer, the use of a variable beamsplitter enables EPR entanglement measurement with any squeezer that would be sufficient for a symmetric two-squeezer experiment.

For comparison, the inseparability for the same biased setup is shown in Figure 12, calculated using the easily measurable expression in Equation 3. Notice that in the absence of asymmetric losses in the entangled beams, the optimal beamsplitter ratio for this measurement is 50:50.

While improving the EPR value is by no means evidence of there being "more" entanglement than there was previously, there are some applications where lowering the EPR value is necessary. For example, the informa-tion transfer rate for QKD systems is strongly dependent on the EPR value of the system.

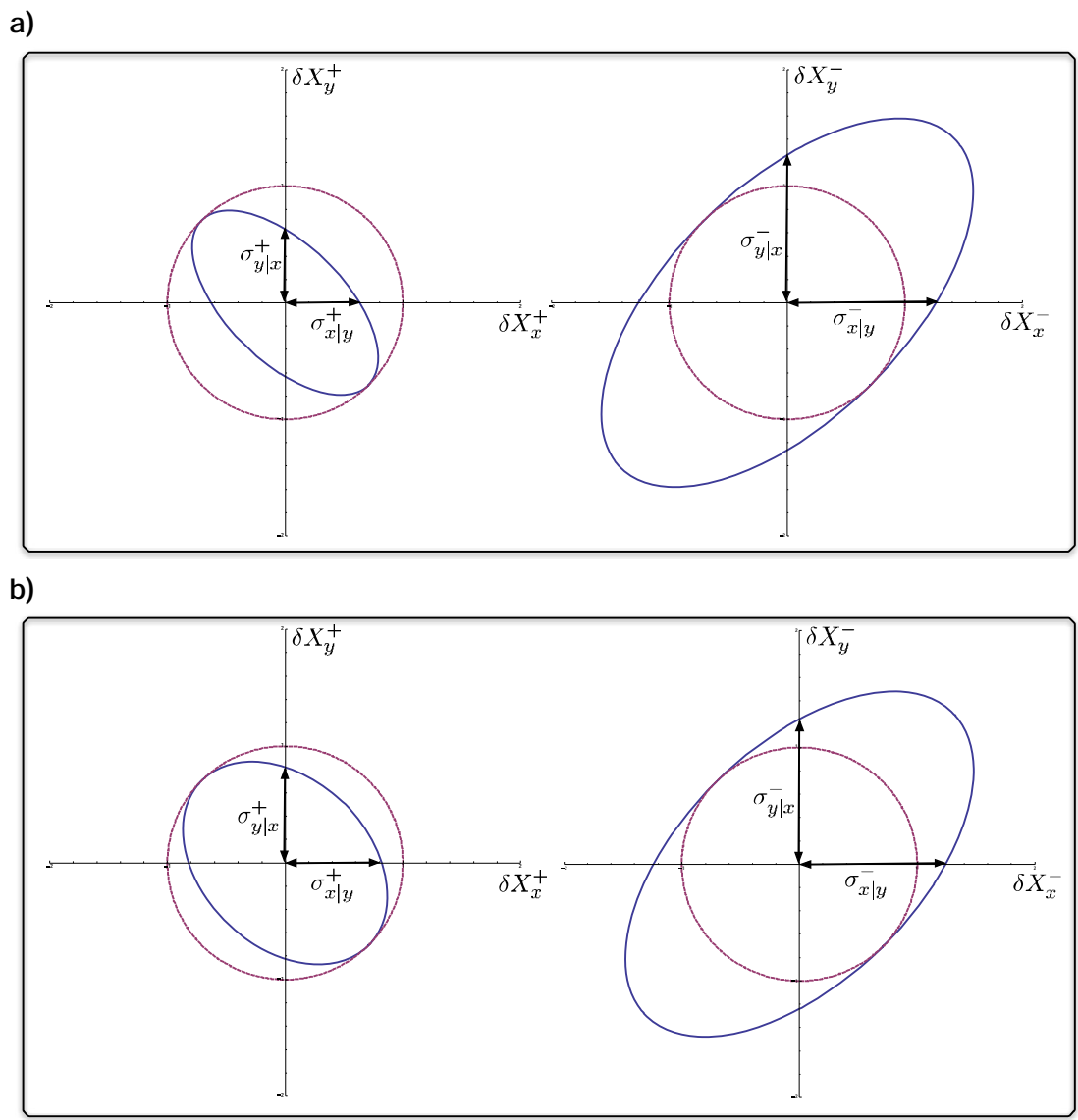

FIG. 8. **a)** The correlation ellipses resulting from a biased entanglement experiment, with no loss on the original squeezed beam. The squeezer used here is -7dB. **b)** A biased entanglement experiment with a loss on the single squeezer.

### B. Experimental Details

An entanglement experiment was performed in order to test the theoretical results that show that EPR measurements can be improved in some instances by changing the beamsplitter ratio. For a given value of loss before the beamsplitter, the beamsplitter ratio was varied, and the EPR value measured at each ratio.

The experimental setup is shown in Figure 13. This diagram shows extra half wave plate/PBS combinations in several beams in case we wished to put extra losses on the beams before or after the main entangling beamsplitter. Ultimately, the original loss in the system was high enough that adding extra losses does not result in EPR entanglement, so these results are not included in this section.

The data was recorded using a National Instruments PXI system. Each detector has two outputs - the high frequency signal (termed AC, or alternating current) and the low frequency signal (termed DC or direct current). The $\delta X_x$ and $\delta X_y$ signals from the subtracted homodyne detectors are sampled at 10MHz, and the data is pro-cessed later.

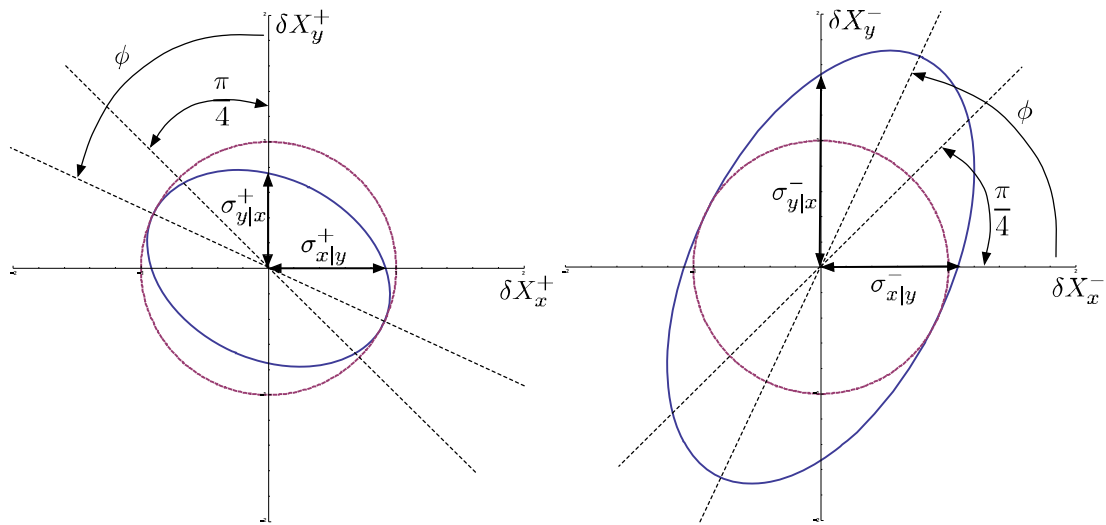

FIG. 9. Biased entanglement experiment with a loss on the original squeezed beam. The beamsplitter used is 80:20 ($t = 0.8$ in Figure 1), and with this ratio EPR entanglement can be observed for $\epsilon_{x|y}$ but not for $\epsilon_{y|x}$.

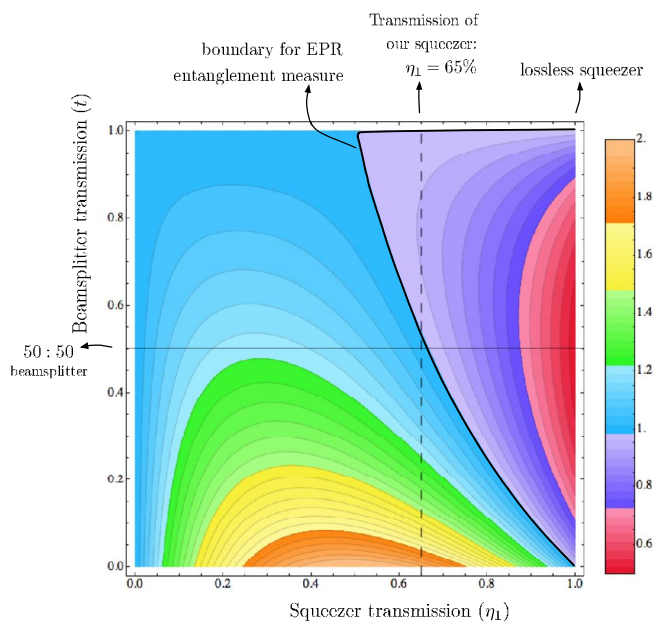

FIG. 10. The theoretical EPR value for a biased entanglement experiment, with a varying value of loss on the squeezer (the transmission $\eta_1$) and a varying beamsplitter transmission, $t$. The usual case for a beamsplitter ratio being 0.5 is marked with the horizontal line, and the transmission of our squeezer at the time of the experiment is marked with the vertical broken line.

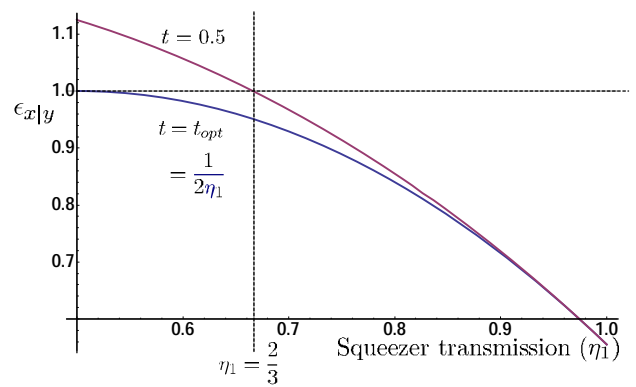

FIG. 11. The theoretical measurement of EPR entanglement in one direction for a setup with a 50:50 beamsplitter ($t = 0.5$), and a setup where the beamsplitter ratio has been optimised according to the loss in the system ($t_{opt} = \frac{1}{2\eta_1}$).

The slow acquisition, at 10kHz, was used to find the amount of power on each of the detectors, which was then used in combination with a Labview program to find the transmission of the entangling beamsplitter in real time.

The data was post processed using Matlab to obtain the variances for the different values that were required. The data was filtered at 4.5 MHz with a frequency window of 1 MHz using a sixth order Butterworth filter. The values of $t$, $V_x$ and $V_y$ for the phase and amplitude quadratures were used to work out the original amount of squeezing as accurately as possible. For each different beamsplitter transmission, these values can be used with Equation 13 to find the original squeezing and antisqueezing. The output of the OPA was 2.9 dB of squeezing and 5.3 dB of antisqueezing. These values were then used to find the theoretical results for the experiment.

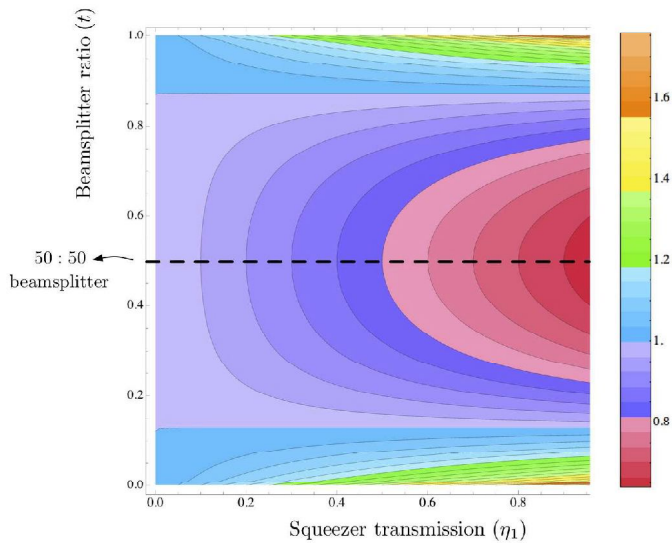

FIG. 12. The theoretical value of inseparability calculated using $\mathcal{I} = \left(V_{x+y}^{\pm} + V_{x-y}^{\mp}\right)/2$. It is assumed that there is one squeezer and no losses after the beamsplitter.

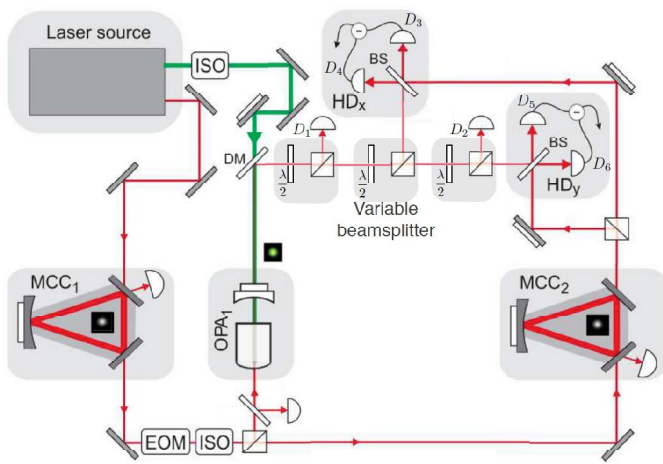

FIG. 13. The experimental setup for the experiment. The experiment had several detectors to record both the instantaneous changes at the homodyne detectors and the DC power measurements for finding the beamsplitter ratio $t$.

Changing the beamsplitter ratio meant that the power in the two entangled beams changed as the experiment was performed. This led to changes in the error signals that then affected the locking systems that were used to lock the phase of the local oscillators to the entangled beams. The locking loops then had to be optimised for each measurement that was made. In some cases, only one of the entangled beams was locked to the required local oscillator phase, and the other local oscillator beam was scanned.

For each beamsplitter ratio used in the experiment, sets of data were taken three or more times. There were two reasons for discarding data that had been taken: for a locked beam, the lock becoming unstable, which can then be seen in the variance trace from the homodyne detector; or, if the local oscillator is being scanned, then the required quadrature might not be reached in some cases, making the data unusable.

Using the same data that yields the variance for each homodyne detector, the conditional variance was found for both quadratures at every beamsplitter ratio investigated.

The results for the EPR entanglement with a varying beamsplitter ratio are shown in Figure 14, alongside the theoretical curve for the experiment. The agreement between the two is within experimental error, and EPR entanglement is achieved with a beamsplitter transmission of 0.78, when it is above one for the case with a 50:50 beamsplitter. At a beamsplitter ratio of 0.78, an EPR value of $0.96 \pm 0.02$ was obtained. The error arises from the changes in the variance due to the locking systems.

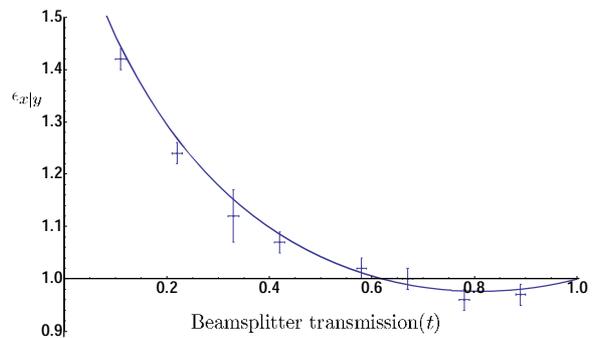

FIG. 14. The results of the biased entanglement experiment, showing the measured EPR entanglement ($\epsilon_{x|y}$) for a varying beamsplitter ratio, $t$. The line is the theoretical curve based on the original level of squeezing (2.9 dB) and antisqueezing (5.3 dB).

## CONCLUSION

We have shown that the direction of the inference that is made to perform an EPR measurement is important in cases where there is an asymmetry in the experiment. It was found that changing the beamsplitter ratio can optimise the measurement made for a lossy system, in some cases enabling an EPR measurement to be made when it would otherwise not be possible. In the case of a biased entanglement setup with just one squeezer, the ideal beamsplitter ratio was found to be $t_{opt} = \frac{1}{2\eta_1}$. This was verified by performing a biased entanglement experiment, with EPR entanglement measured in cases where an apparatus with a 50:50 beamsplitter yields only classical correlations.

These calculations can easily be extended to N-mode entanglement, rather than the 2-mode calculations used here. This then allows such experiments for the generation of multimode entangled states to be planned so that the resources are distributed in a way that is optimised for the measurement being made.

## ACKNOWLEDGMENTS

This research was conducted by the Australian Research Council Centre of Excellence for Quantum-Atom Optics. The authors would like to acknowledge the contribution of the HIDEAS FP7 research program.